\documentstyle[12pt,psfig]{article}

\newcommand{\gsim}{\;\rlap{\lower 3.5 pt \hbox{$\mathchar \sim$}} \raise 1pt
 \hbox {$>$}\;}
\newcommand{\lsim}{\;\rlap{\lower 3.5 pt \hbox{$\mathchar \sim$}} \raise 1pt
 \hbox {$<$}\;}
\newcommand{\di}{\mathop{\mbox{Li}_2}\nolimits}
\newcommand{\re}{\mathop{\mbox{Re}}\nolimits}
\newcommand{\msbar}{\overline{\rm MS}}

\begin{document}

\title{\vskip-3cm{\baselineskip14pt
\centerline{\normalsize\hfill PM 97/45}
\centerline{\normalsize\hfill MPI/PhT/97--71}
\centerline{\normalsize\hfill TUM--HEP--301/97}
\centerline{\normalsize\hfill hep--ph/9712330}
\centerline{\normalsize\hfill December 1997}
}
\vskip1.5cm
Two-Loop Electroweak Heavy-Fermion Corrections to Higgs-Boson Production and 
Decay}
\author{{\sc A. Djouadi$^{1,2}$, P. Gambino$^3$ and B.A. Kniehl$^2$}\\
{\normalsize $^1$ Laboratoire de Physique Math\'ematique,
Universit\'e de Montpellier~II,}\\
{\normalsize \phantom{$^1$} Place Eug\`ene Bataillon, 34095 Montpellier,
France}\\
{\normalsize $^2$ Max-Planck-Institut f\"ur Physik 
(Werner-Heisenberg-Institut),}\\
{\normalsize \phantom{$^3$} F\"ohringer Ring~6, 80805 Munich, Germany} \\
{\normalsize $^3$ Institut f\"ur Theoretische Physik,
Technische Universit\"at M\"unchen,}\\
{\normalsize \phantom{$^3$} James-Franck-Stra\ss e, 85747 Garching, Germany}
}
\date{}
\maketitle
\begin{abstract}
The dominant electroweak corrections to the production cross sections and
partial decay widths of a light Standard-Model Higgs boson, with mass
$M_H\ll2m_t$, are due to top-quark loops.
By means of a low-energy theorem, we study at the two-loop level the leading
shifts in the Higgs-boson couplings to pairs of light fermions and gauge
bosons induced by a sequential isodoublet of high-mass fermions.
For tree-level and loop-induced Higgs-boson couplings, these corrections are
of relative orders ${\cal O}(G_\mu^2 m_F^4)$ and ${\cal O}(G_\mu m_F^2)$,
respectively, where $m_F$ represents a generic heavy-fermion mass, with
$m_F\gg M_W,M_H$.
Except for the $Hb\bar b$ coupling, all results carry over to the case of the
top-quark-induced corrections.
We discuss possible phenomenological consequences of our results.

\end{abstract}

\newpage

\section{Introduction}

The Higgs boson is the last missing link of the Standard Model (SM) of
elementary particle physics. Its experimental discovery would eventually solve
the longstanding puzzle as to whether nature makes use of the mechanism of
spontaneous symmetry breaking to generate the particle masses. So far, direct
searches at the CERN Large Electron Positron Collider (LEP) have been able to
rule out the mass range $M_H\le77.1$~GeV at the 95\% confidence level (CL) 
\cite{jan}. On the other hand, exploiting the sensitivity to the Higgs boson
via quantum loops, a global fit to the latest electroweak precision data
predicts $M_H=115{+116\atop-66}$GeV together with a 95\% CL upper
bound at 420~GeV \cite{cla}; see also the discussion in Ref.~\cite{DGPS}.

The study of quantum corrections to the production and decay processes of the
Higgs boson has received much attention in the literature; for a review, see
Ref.~\cite{kni}. In the following, we shall consider a light-Higgs-boson
scenario,  characterized by $M_H\ll2m_t$. Such a situation is not only
favored experimentally \cite{cla,DGPS}, but also from various theoretical
considerations \cite{cab,cap}. For example, the requirement that the running
Higgs self-coupling, $\lambda(\mu)$, does not develop a Landau pole for
renormalization scales $\mu<\Lambda_{\rm GUT}\approx10^{16}$~GeV leads to the
triviality bound $M_H\lsim185$~GeV \cite{cab}. Since the top quark, with pole
mass $m_t=(175.6\pm5.5)$~GeV \cite{abe}, is much heavier than the intermediate
bosons, top-quark loop effects tend to be the dominant source of electroweak
corrections to the couplings of a light Higgs boson.
The goal of this paper is to calculate the leading $m_t$-dependent
corrections at two loops to the Higgs-boson couplings to gauge bosons and
light fermions.
For the time being, we shall leave aside the bottom and top Yukawa couplings
as well as the Higgs-boson self-couplings.

For the discussion of the leading $m_t$-dependent corrections, it is useful to
distinguish between the Higgs-boson couplings that already exist at tree
level, such as the $Hf\bar f$, $HWW$ and $HZZ$ couplings, and those which are
induced by heavy-particle loops, such as the $H\gamma\gamma$, $H\gamma Z$ and
$Hgg$ couplings.
The present knowledge of the higher-order radiative corrections to the
couplings of the first type is as follows.
The complete one-loop calculation of the $Hf\bar f$ and $HVV$ couplings, where
$V=W,Z$, has been carried out in Refs.~\cite{cha,hff} and \cite{cha,hvv},
respectively.
The two-loop ${\cal O}(\alpha_s G_\mu m_t^2)$ corrections to the light-fermion
and bottom Yukawa couplings have been calculated in Refs.~\cite{alb,adj} and
\cite{hbb}, respectively, and those to the $HVV$ couplings in Ref.~\cite{sea}.
The three-loop ${\cal O}(\alpha_s^2 G_\mu m_t^2)$ corrections
may be found in Ref.~\cite{mat} for the $H\l^+ l^-$ and $HVV$
couplings and in Ref.~\cite{kos} for the $Hq\bar q$ couplings, including the
$b\bar b$ case. As for the loop-induced Higgs-boson vertices, the
two-loop QCD corrections to the $H\gamma\gamma$, $H\gamma Z$ and $Hgg$
couplings have been derived in Refs.~\cite{zhe,spi,ina}, respectively.
The three-loop ${\cal O}(\alpha_s^2)$ corrections  to the $H\gamma\gamma$ and
$Hgg$ and couplings have been found in Refs.~\cite{ste,dec} and \cite{hgg},
respectively; this analysis has also been extended to ${\cal O}(\alpha_s^3)$
\cite{dec}.
The leading two-loop electroweak corrections, which are of
${\cal O}(G_\mu m_t^2)$ relative to the one-loop result, 
have only been calculated in the case of the $Hgg$ coupling \cite{djo}.

The leading electroweak two-loop corrections of ${\cal O}(G_\mu^2 m_t^4)$ to
the $Hf\bar f$ and $HVV$ couplings, as well as the two-loop
${\cal O}(G_\mu m_t^2)$ corrections to the $H\gamma\gamma$ and $H\gamma Z$
couplings are not yet available.
In this paper, we shall partly fill these gaps.
For the sake of generality, we shall consider a sequential fourth-generation
fermion doublet with arbitrary masses $m_A,m_B\gg M_W,M_H$.
The results for the top-quark-induced corrections then emerge by substituting
$m_A=m_t$ and $m_B=0$.
Notice, however, that there is an additional class of diagrams contributing to
the ${\cal O}(G_\mu^2 m_t^4)$ corrections to the $Hb\bar b$  coupling, which
is absent in the fourth-generation case.
These bottom-specific corrections will not be considered here.

The study of the extension of the SM by a fourth fermion
generation is interesting in its own right. Recently, it was noticed \cite{cel}
that arguments favoring the presence of a  fourth fermion generation may be
adduced on the basis of the democratic mass-matrix approach \cite{har}. The
possible existence of a fourth fermion generation is also considered in the
Review of Particle Physics \cite{pdg}, where mass bounds are listed. For a
recent model-independent analysis, see Ref.~\cite{nov}. It is advantageous to
trace such novel fermions via their loop effects in the Higgs sector, since
these effects are also sensitive to mass-degenerate isodoublets via
fermion-mass power corrections. This has originally been observed in
Ref.~\cite{cha} in connection with the $Hf\bar f$, $HWW$ and $HZZ$
couplings.
Moreover, the $H\gamma\gamma$ and $Hgg$ couplings may serve as devices to
detect mass-degenerate isodoublets of ultraheavy fermions \cite{wil}, although
power corrections do not occur here at leading order.
By contrast, in the gauge sector, leading power corrections only appear in
connection with isospin breaking \cite{ros}.

Since, at this point, we only wish to extract the leading heavy-fermion
corrections to the Higgs-boson couplings, we may work in the framework of a
Yukawa Lagrangian where the heavy fermions only couple to the Higgs boson and
the longitudinal components of the gauge bosons, a situation which
corresponds to the {\it gaugeless limit} of the SM \cite{bar}.
In the presence of external gauge bosons, these can be considered as
background sources.
In practice, in the covariant gauges, it suffices to only consider the
diagrams that involve exchanges of scalar bosons.
However, this simple approach may fail if there is a symmetry which lowers the
superficial degree of divergence of the loops.
For instance, this happens in the case of the $H\gamma\gamma$
coupling, where QED Ward identities imply that $W$-boson loops cannot be
neglected any more. Nevertheless, we will show that, also in this case, the
calculation can be performed in a very compact manner.
Moreover, we may also take advantage of a soft-Higgs theorem
\cite{let1,russians}, extended to higher-orders \cite{let2}, which relates the
couplings of the Higgs boson to gauge bosons to derivatives of the
vacuum-polarization functions of the latter.

This paper is organized as follows. In the next section, we will analyze the
universal ${\cal O}(G_\mu^2 m_F^4)$ correction to the Higgs-boson vertices.
This is the only such contribution in the case of the lepton and light-quark
Yukawa couplings, except for the $Hb\bar b$ coupling.
The results will be presented both in the electroweak $\msbar$ \cite{msbar}
and on-shell \cite{si80,renorm} schemes. In Section~3, we will
derive the ${\cal O}(G_\mu^2 m_F^4)$ corrections to the $HWW$ and $HZZ$
couplings.
In Section~4, we will turn to the couplings which are not present at tree
level, and calculate the two-loop ${\cal O}( G_\mu m_F^2)$ corrections to the 
loop-induced $H\gamma\gamma$ and $Hgg$ couplings.
Finally, Section~5 is devoted to the numerical discussion of our results.

\boldmath
\section{Universal correction: $Hf\bar f$ couplings}
\unboldmath

At leading order in the heavy-fermion masses, i.e.\ to ${\cal O}(G_\mu m_F^2)$
at one loop and to ${\cal O}(G_\mu^2 m_F^4)$ at two loops for couplings present 
at the tree level, one can isolate an ultraviolet-finite and gauge-invariant
contribution to the Higgs-boson production and decay amplitudes which is
universal in the sense that it is independent of the initial and final states.
As for the Yukawa couplings of the leptons and light quarks, except for the
bottom quark, this is the only source of leading top-quark contributions.
In this section, we will discuss this universal correction in the more general
case of an isodoublet of heavy fermions, first in the electroweak $\msbar$
scheme of Ref.~\cite{msbar} and then in the on-shell scheme.
The scheme of Ref.~\cite{msbar} uses $\msbar$ couplings and on-shell masses as 
basic parameters.

The bare Lagrangian describing the interactions of the Higgs boson with
fermions $f$ and intermediate bosons $V=W,Z$,
\begin{equation}
{\cal L}=\frac{H_0}{v_0}\left(-\sum_fm_{f0}\bar f_0f_0
+2M_{W0}^2W_{0\mu}^\dag W_0^\mu+M_{Z0}^2Z_{0\mu}Z_0^\mu\right), 
\end{equation}
contains the overall factor $H_0/v_0$, where $H$ the Higgs field, $v$ is the
vacuum expectation value and the subscript $0$ labels bare quantities.
This ratio undergoes a finite renormalization.
Recalling that $v_0=2M_{W0}/g_0=2c_0M_{Z0}/g_0$, where $g$ is the SU(2)$_L$
coupling constant and $c$ is the cosine of the weak mixing angle, we define
the finite universal shift in the $\msbar$ scheme, $\hat\delta_u$, via
\begin{equation}
\frac{H_0}{v_0}=\frac{\hat gH}{2\hat cM_Z}\left(1+\hat\delta_u\right),
\label{univ}
\end{equation}
where $\hat g$ and $\hat c$ are defined in the $\msbar$ scheme and $M_Z$ is 
the physical $Z$-boson mass defined as the real part of the pole of the
propagator.
Shifting $M_{Z0}^2$ and $H_0$, one obtains
\begin{equation}
\frac{H_0}{v_0}=\frac{\hat gH}{2\hat cM_Z} 
\left(1-\frac{\delta M_Z^2}{M_Z^2}\right)^{-1/2}
\left[1+\re\Pi_{HH}^\prime(M_H^2)\right]^{-1/2},
\label{uni}
\end{equation}
where $\delta M_Z^2=\re A_{ZZ}(M_Z^2)$, $A_{ZZ}(q^2)$ is the $Z$-boson
self-energy and $\Pi_{HH}(q^2)$ is the Higgs-boson self-energy, defined
according to the conventions of Ref.~\cite{alb,adj}.
Since we are only interested in the leading ${\cal O}(G_\mu m_F^2)$ and
${\cal O}(G_\mu^2 m_F^4)$ corrections, we may replace $\re A_{ZZ}(M_Z^2)$ and
$\re\Pi_{HH}^\prime(M_H^2)$ in Eq.~(\ref{uni}) with $A_{ZZ}(0)$ and
$\Pi_{HH}^\prime(0)$, respectively.
In the $\msbar$ scheme, as a consequence of the SU(2) custodial symmetry of
the SM, the counterterm
$\delta\hat c^2=-\hat c^2[\delta M_W^2/M_W^2-\delta M_Z^2/M_Z^2]_{\rm UV}$,
where the subscript $UV$ indicates that only the
ultraviolet pole has to be considered, receives no quadratic fermion-mass
contributions \cite{msbar}.
Hence, the counterterms for $\hat{c}$ and $\hat{g}$
do not contain leading-order contributions in the heavy-fermion masses.
This is the reason why we first present our results in this particular 
scheme. Denoting one- and two-loop contributions by the superscripts (1) and
(2), respectively, we may identify the ${\cal O}(G_\mu m_F^2)$ contribution to
$\hat\delta_u$ as
\begin{equation}
\hat\delta_u^{(1)}=\frac{1}{2}\,\frac{\delta^{(1)}M_Z^2}{M_Z^2}
-\frac{1}{2}\Pi_{HH}^{(1)\prime}(0).
\label{univ1}
\end{equation}
The one-loop ${\cal O}(G_\mu m_F^2)$ contribution from a weak isodoublet of
ultraheavy fermions, with arbitrary masses $m_A$, $m_B$ and color multiplicity
$N_c$, to the $Z$-boson mass counterterm and the Higgs-boson wave-function
renormalization constant (see Fig.~1a) are given by
\begin{eqnarray}
\frac{\delta^{(1)}M_Z^2}{M_Z^2}&=&\frac{N_cg_0^2}{16\pi^2c_0^2M_Z^2}
\sum_{F=A,B}m_{F0}^2\left(\frac{\mu}{m_{F0}}\right)^{2\epsilon}
\left[\frac{1}{2\epsilon}+\frac{\pi^2}{24}\epsilon+{\cal O}(\epsilon^2)
\right],
\label{azz1}\\
\Pi_{HH}^{(1)\prime}(0)&=&\frac{N_cg_0^2}{16\pi^2c_0^2 M_{Z0}^2}
\sum_{F=A,B}m_{F0}^2\left(\frac{\mu}{m_{F0}}\right)^{2\epsilon}
\left[\frac{1}{2\epsilon}-\frac{1}{3}+\frac{\pi^2}{24}\epsilon
+{\cal O}(\epsilon^2)\right],
\label{pihh1}
\end{eqnarray}
respectively.
Here, $n=4-2\epsilon$ is the space-time dimension in dimensional
regularization, $\mu$ the 't~Hooft renormalization scale and we have kept 
the ${\cal O}(\epsilon)$ terms for later convenience.
Inserting Eqs.~(\ref{azz1},\ref{pihh1}) into Eq.~(\ref{univ1}), one obtains
\begin{equation}
\hat\delta_u^{(1)}=\frac{N_c\hat g^2m_A^2}{16\pi^2\hat c^2M_Z^2}\,
\frac{1+x}{6}.
\label{ms1}
\end{equation}
Here and in the following, $x=m_B^2/m_A^2$.

We now extend the analysis to the two-loop level keeping only the 
leading ${\cal O}(G_\mu^2 m_F^4)$ terms.
Our starting point is the two-loop analogue of Eq.~(\ref{univ1}),
\begin{equation}
\hat\delta_u^{(2)}=\frac{1}{2}\,\frac{\delta^{(2)}M_Z^2}{M_Z^2}
-\frac{1}{2}\Pi_{HH}^{(2)\prime}(0)+\delta_u^{\rm ct}
+\frac{1}{2}\,\frac{\delta^{(1)}M_Z^2}{M_Z^2}\Pi_{HH}^{(1)\prime}(0)
+\frac{3}{2}\left(\hat{\delta}_u^{(1)}\right)^2.
\label{univ2}
\end{equation}
Here, the first two terms represent the irreducible two-loop contributions 
to the $Z$-boson mass counterterm and the Higgs-boson wave-function
renormalization constant to be calculated below.
The third term arises because part of the one-loop contribution of
Eq.~(\ref{ms1}) is expressed in terms of bare quantities, namely the fermion
masses in Eqs.~(\ref{azz1},\ref{pihh1})
and the $Z$-boson mass in Eq.~(\ref{pihh1}), while the one appearing
in Eq.~(\ref{azz1}) is renormalized by definition. The 
counterterms associated with the couplings can be neglected, as 
they do not contain leading terms in the $\msbar$ scheme. The last two terms 
in Eq.~(\ref{univ2}) represent iterations of the one-loop contributions.

At this point, it is useful to separate $\hat\delta_u^{(2)}$ in two parts:
one which is proportional to $N_c$ and includes all irreducible two-loop
contributions along with the respective fermion mass counterterms; and one
which is proportional to $N_c^2$ and comes from the iteration of the one-loop
terms including the $Z$-boson mass counterterms.
The two sets of contributions must be independently finite and gauge invariant.
As the calculation is performed in $n$ dimensions, ${\cal O}(\epsilon)$ terms
must generally be kept in the one-loop expressions.
We first consider the subset of contributions proportional to $N_c^2$.
Shifting the bare $Z$-boson mass in Eq.~(\ref{pihh1}), one obtains a term which 
exactly cancels the fourth term in Eq.~(\ref{univ2}).
Thus, the only contribution proportional to $N_c^2$ arises from the last term
in Eq.~(\ref{univ2}), with $\hat{\delta}_u^{(1)}$ given by Eq.~(\ref{ms1}).
 
As for the contributions proportional to $N_c$, let us first consider the 
heavy-fermion mass counterterms (see Fig.~1b).
Since $m_A$ and $m_B$ only enter at one loop, it is sufficient to know these 
counterterms to ${\cal O}(G_\mu m_F^2)$.
If the fermion masses are renormalized on shell, we have
\begin{eqnarray}
\frac{\delta m_A}{m_A}&=&\frac{\hat g^2m_A^2}{16\pi^2\hat c^2M_Z^2}
\left(\frac{\mu}{m_A}\right)^{2\epsilon}
\left[-{3(1-x)\over8\epsilon}-\frac{8-5x+x^2}{8}-\frac{x^2}{8}(3-x)\ln x
\right.\nonumber\\
&&+\left.\frac{(1-x)^3}{8}\ln|1-x|+{\cal O}(\epsilon)
\right].
\label{quarkct}
\end{eqnarray}
The corresponding expression for $\delta m_B/m_B$ emerges by interchanging
$A\leftrightarrow B$.
Since $\hat\delta_u^{(1)}$ in Eq.~(\ref{ms1}) is a finite quantity, we do not
need the ${\cal O}(\epsilon)$ parts of the fermion mass counterterms.
However, it is necessary to retain the ${\cal O}(\epsilon)$ piece in the bare
analogue of Eq.~(\ref{ms1}), since it combines with ${\cal O}(1/\epsilon)$
terms of the bare fermion masses contained therein to give finite two-loop
contributions.
With the help of Eqs.~(\ref{azz1},\ref{pihh1},\ref{quarkct}), we thus find
\begin{equation}
\delta_u^{\rm ct}\doteq{(1-x)^2\over8\epsilon}+{3-2x+3x^2\over12} 
+{x\over24}(3-2x-x^2)\ln x+{(1-x)^4\over24x}\ln|1-x|. 
\label{ctNc}
\end{equation}
Throughout Sections~2 and 3, the symbol $\doteq$ indicates that the right-hand
side is to be multiplied by the overall factor
\begin{equation}
\frac{N_c\hat g^4m_A^4}{(16\pi^2)^2\hat c^4M_Z^4}
\left(\frac{\mu}{m_A}\right)^{4\epsilon},
\label{factor}
\end{equation}
where $\epsilon=0$ if finite results are concerned.

We now consider the irreducible two-loop contributions to
$\hat\delta_u^{(2)}$, i.e.\ the first two terms of Eq.~(\ref{univ2}), which
all scale like $N_c$.
Typical Feynman diagrams are depicted in Fig.~2.
We work in the $R_\xi$ gauge and take $\gamma_5$ to be anticommuting.
Since we are only interested in the leading ${\cal O}(G_\mu^2 m_F^4)$ terms,
we may set the Higgs- and intermediate-boson masses as well as the external
momentum squared to zero from the beginning.
The actual calculation is conveniently performed with the aid of the symbolic
manipulation program package {\it ProcessDiagram} \cite{program}.
Since all trilinear scalar-boson vertices, which are proportional to the 
scalar-boson masses, are set to zero from the beginning, only the diagrams
with four Yukawa couplings contribute to the Higgs-boson self-energy in
leading order (see Fig.~2a). The Higgs-boson wave-function renormalization
constant is found to be
\begin{eqnarray}
\Pi_{HH}^{(2)\prime}(0)&\doteq&-{3(1-x)^2\over16\epsilon^2}
+\frac{1}{\epsilon}\left[{11-38x+11x^2\over32}-{3\over8}x(1-x)\ln x\right]
\nonumber\\
&&+{17-2x+17x^2\over192}-{\pi^2\over32}(1-x)^2+{x\over16}(19-11x)\ln x
\nonumber\\
&&+{x\over16}(3-5x)\ln^2x-{(1-x^2)\over 8}\di(1-x)+{\cal O}(\epsilon),
\label{pihh2}
\end{eqnarray}
where $\di(x)=-\int_0^1dt\ln(1-xt)/t$ is the Spence function. 

In the limit we are considering, the $Z$-boson mass counterterm only receives
contributions from diagrams which contain no internal vector-boson lines and
at least one scalar-boson line (see Fig.~2b). Some of these diagrams are
infrared divergent because we have set the masses of the Higgs and Goldstone
bosons to zero. One can regulate these divergences in different
ways, for instance by using dimensional regularization or by assigning small
masses to these scalar bosons, which may be different or the same.
Note that, in the Lagrangian under consideration, there is no symmetry which
protects the masslessness of the scalar bosons, so that these masses receive
radiative corrections, which are also of ${\cal O}(G_\mu^2m_F^4)$.
Since these contributions are linked to the tadpole counterterms via Ward
identities, as explained e.g.\ in the Appendix of Ref.~\cite{alb}, we can
equivalently view them as keeping the Goldstone bosons massless.
Therefore, we still need to add the counterterm contributions which emerge
by shifting the scalar-bosons masses in the one-loop seed diagrams.
These contributions are themselves infrared divergent, but if they are 
combined with the irreducible two-loop diagrams, the result is infrared
regular and independent of the regularization adopted, as it should be.
The use of a small mass regulator common to all the scalar bosons
is particularly convenient, since it quenches the counterterm contribution.
In this way, we find for $\delta^{(2)}M_Z^2/M_Z^2$, including the scalar-boson
mass-counterterm contributions,
\begin{eqnarray}
\frac{\delta^{(2)}M_Z^2}{M_Z^2}&\doteq&-{3(1-x)^2\over16\epsilon^2}
+\frac{1}{\epsilon}\left[{3-22x+3x^2\over32}-\frac{3}{8}x(1-x)\ln x\right]
\nonumber\\ 
&&+{19\over64}(1-x)^2-{\pi^2\over32}(1-x)^2+{x\over16}(11-3x)\ln x
\nonumber\\
&&+{x\over16}(3-5x)\ln^2x-{1-x^2\over8}\di(1-x)+{\cal O}(\epsilon).
\label{az2}
\end{eqnarray}

Inserting Eqs.~(\ref{ms1},\ref{ctNc},\ref{pihh2},\ref{az2}) into
Eq.~(\ref{univ2}), we finally obtain the universal correction at two loops in
the $\msbar$ scheme as
\begin{equation}
\hat\delta_u^{(2)}\doteq{17-22x+17x^2\over48}-{x\over24}(3-4x+x^2)\ln x
+{(1-x)^4\over24x}\ln|1-x|+\frac{N_c}{24}(1+x)^2.
\label{univ2a}
\end{equation}
In the limiting cases $m_A\gg m_B$ and $m_A\approx m_B$, this becomes
\begin{equation}
\left.\hat\delta_u^{(2)}\right|_{x=0}\doteq\frac{5}{16}+\frac{N_c}{24},
\qquad
\left.\hat\delta_u^{(2)}\right|_{x=1}\doteq\frac{1}{4}+\frac{N_c}{6},
\end{equation}
respectively.

In the next section, we will also need the leading contribution from an
ultraheavy fermion isodoublet to the $W$-boson mass counterterm
$\delta M_W^2=\re A_{WW}(M_W^2)$, which may be calculated in the same way as
in the $Z$-boson case.
At the one-loop order, it is given by 
\begin{eqnarray}
\frac{\delta^{(1)}M_W^2}{M_W^2}&=&\frac{N_cg_0^2m_A^2}{16\pi^2M_W^2}
\left(\frac{\mu}{m_A}\right)^{2\epsilon}
\left\{{1+x\over2\epsilon}+{1+x\over4}+{x^2\over2(1-x)}\ln x\right.\nonumber\\
&&+\left.\epsilon\left[{1+x\over8}+{\pi^2\over24}(1+x)+{x^2\over4(1-x)}\ln x
-{x^2\over4(1-x)}\ln^2x\right]\right\}+{\cal O}(\epsilon^2).
\quad\label{aww1}
\end{eqnarray}
The corresponding two-loop result, including the contributions due to the
scalar-boson mass counterterms, is given by
\begin{eqnarray}
\frac{\delta^{(2)}M_W^2}{M_W^2}&\doteq&-{3(1-x)^2\over16\epsilon^2}
+\frac{1}{\epsilon}\left[-{3-14x+3x^2\over32}+{3x^2(3-x)\over8(1-x)}\ln x
\right]\nonumber\\
&&+{69+14x+69x^2\over64}-{\pi^2\over32}(1-x)^2
+{x(14+31x-3x^2)\over16(1-x)}\ln x
\nonumber\\
&&-{x^2(7-10x-x^2)\over16(1-x)^2}\ln^2x-{7\over8}(1-x^2)\di(1-x) 
+{\cal O}(\epsilon).
\label{aw2}
\end{eqnarray}

\subsection*{Results in the on-shell scheme}

So far, we have discussed the universal correction in the electroweak $\msbar$
scheme of Ref.~\cite{msbar}.
By contrast, in the electroweak on-shell scheme, the couplings are expressed
in terms of $G_\mu$, $M_W$ and $M_Z$.
To the orders considered here, the relations between the $\msbar$ couplings
and on-shell parameters read \cite{msbar} 
\begin{equation}
\hat c^2=\frac{M_W^2}{M_Z^2}(1-\Delta\rho),\qquad
\hat g^2=4\sqrt2G_\mu M_W^2,
\label{replace}
\end{equation}
where
\begin{equation}
\Delta\rho=\frac{A_{ZZ}(0)}{M_Z^2}-\frac{A_{WW}(0)}{M_W^2}
\end{equation}
embodies the leading fermionic correction to the electroweak $\rho$ parameter.
The one-loop heavy-fermion contribution to $\Delta\rho$ is well known and
reads \cite{ros}
\begin{equation}
\Delta\rho^{(1)}=\frac{N_cG_\mu m_A^2}{8\pi^2\sqrt2}
\left(1+x+\frac{2x}{1-x}\ln x\right).
\label{drho1}
\end{equation}
We recall that this expression vanishes for degenerate fermion masses, as 
isospin violation is the only source of custodial symmetry breaking, which
induces leading power corrections.
On the other hand, $\Delta\rho^{(1)}=N_cG_\mu m_A^2/\left(8\pi^2\sqrt2\right)$
if $m_A\gg m_B$.
The two-loop contribution is obtained from the difference of
Eqs.~(\ref{az2},\ref{aw2}) as
\begin{eqnarray}
\Delta\rho^{(2)}&\doteq&{11-12x+11x^2\over8}
+{2+x+19x^2-9x^3+3x^4\over16(1-x)}\ln x
\nonumber\\
&&+{(1-x)^2\over16x}(3-2x+3x^2)\ln|1-x|-{x^3(3-x^2)\over8(1-x)^2}\ln^2x
\nonumber\\
&&+{1-x\over8x}(1+x^3)\ln x\ln|1-x|-\frac{3}{4}(1-x^2)\di(1-x),
\label{drho2}
\end{eqnarray}
which is in agreement with Ref.~\cite{hoog}.
Again, the expression vanishes for $m_A=m_B$.
On the other hand, for $m_A\gg m_B$, one has
$\Delta\rho^{(2)}\doteq(19-2\pi^2)/16$.

In order to derive the universal contribution in the on-shell scheme,
$\delta_u$, we start from the identity
\begin{equation}
\frac{H_0}{v_0}=\frac{\hat gH}{2\hat cM_Z}\left(1+\hat\delta_u\right)
=2^{1/4}G_\mu^{1/2}H(1+\delta_u).
\label{univ3}
\end{equation}
Using Eq.~(\ref{replace}), we thus have
\begin{equation}
1+\delta_u=\left(1+\hat\delta_u\right)(1-\Delta\rho)^{-1/2}. 
\label{conversion}
\end{equation}
At the one loop-level, this implies
\begin{equation}
\delta_u^{(1)}=\frac{N_cG_\mu m_A^2}{8\pi^2\sqrt2}
\left[\frac{7}{6}(1+x)+\frac{x}{1-x}\ln x\right],
\label{os1}
\end{equation}
which agrees with Refs.~\cite{cha,hff}.
Expanding Eq.~(\ref{conversion}) consistently through two loops, we obtain
\begin{equation}
\delta_u^{(2)}=\hat\delta_u^{(2)}+\frac{1}{2}\Delta\rho^{(2)}
+\frac{1}{2}\hat\delta_u^{(1)}\Delta\rho^{(1)}
+\frac{3}{8}\left(\Delta\rho^{(1)}\right)^2.
\end{equation}
Inserting Eqs.~(\ref{univ1},\ref{univ2a},\ref{drho1},\ref{drho2}), one then
finds the universal correction in the electroweak on-shell scheme to be
\begin{eqnarray}
\delta_u^{(2)}&\doteq&
\frac{25-29x+25x^2}{24}+\frac{6-9x+85x^2-47x^3+13x^4}{96(1-x)}\ln x
\nonumber\\
&&+\frac{(1-x)^2}{96x}(13-14x+13x^2)\ln|1-x|
-\frac{x^3(3-x^2)}{16(1-x)^2}\ln^2x
\nonumber\\
&&+\frac{1-x}{16x}(1+x^3)\ln x\ln|1-x|-\frac{3}{8}(1-x^2)\di(1-x)
\nonumber\\
&&+N_c\left[\frac{11}{128}(1+x)^2+\frac{13x(1+x)}{96(1-x)}\ln x
+\frac{3x^2}{32(1-x)^2}\ln^2x\right],
\end{eqnarray}
which, for $m_A\gg m_B$ and $m_A\approx m_B$, takes the values
\begin{equation}
\left.\delta_u^{(2)}\right|_{x=0}\doteq\frac{29}{32}-\frac{\pi^2}{16}
+\frac{11}{128}N_c,\qquad
\left.\delta_u^{(2)}\right|_{x=1}\doteq\frac{1}{4}+\frac{N_c}{6},
\end{equation}
respectively.
This completes the discussion of the universal contribution and the 
leading corrections to the lepton and light-quark Yukawa couplings.

\boldmath
\section{The $HWW$ and $HZZ$ couplings}
\unboldmath

We now turn to the leading electroweak corrections to couplings between Higgs
and intermediate bosons.
Since the $HWW$ and $HZZ$ couplings appear already at the tree level, the
leading one- and two-loop corrections in the heavy-fermion masses are of
${\cal O}(G_\mu m_F^2)$ and ${\cal O}(G_\mu^2 m_F^4)$, respectively.

In the electroweak $\msbar$ scheme, the renormalized $HVV$ interaction
Lagrangian takes the form
\begin{equation}
{\cal L}_{HVV}=\frac{\hat gH}{\hat cM_Z}\left(1+\hat\delta_u\right)
\left[M_W^2 W_\mu^\dag W^\mu\left(1+\hat\delta_W\right)
+\frac{M_Z^2}{2}Z_\mu Z^\mu\left(1+\hat\delta_Z\right)\right],
\label{lagrHVV}
\end{equation}
where $\hat\delta_u$ is given by Eq.~(\ref{univ2a}) and
\begin{equation}
1+\hat\delta_V=\left(1+\delta_V^{\rm irr}\right)
\left(1-\frac{\delta M_V^2}{M_V^2}\right).
\label{delv}
\end{equation}
The appropriate one- and two-loop expressions for $\delta M_V^2/M_V^2$ may be
found in Eqs.~(\ref{azz1},\ref{az2}, \ref{aww1},\ref{aw2}).
Notice that the wave-function renormalization constants of the intermediate
bosons do not contribute to leading order in the heavy-fermion masses.
The only missing piece in Eq.~(\ref{lagrHVV}) is the irreducible $HVV$ vertex
correction, $\delta_V^{\rm irr}$.
It can be derived by means of a low-energy theorem which relates the
heavy-fermion contributions to the $HVV$ vertex and $VV$ self-energy
corrections in the following way:
\begin{equation}
\delta_V^{\rm irr}=\sum_{F=A,B}\frac{m_{F0}^2\partial}{\partial m_{F0}^2}\,
\frac{A_{VV}(0)}{M_{V0}^2},
\label{let}
\end{equation}
where it is understood that the masses appearing in coupling constants must be
treated as constants under the differentiation \cite{let2}.

In terms of bare parameters, the one-loop results for $\delta_V^{\rm irr}$
through ${\cal O}(\epsilon)$ read
\begin{eqnarray}
\delta_W^{\rm irr(1)}&=&\frac{N_cg_0^2m_{A0}^2}{16\pi^2M_{W0}^2}
\left(\frac{\mu}{m_{A0}}\right)^{2\epsilon}
\left\{{1+x_0\over2\epsilon}-{1+x_0\over4}+{x_0^2\over2(1-x_0)}\ln x_0 
\right.\nonumber\\
&&+\left.\epsilon\left[-\frac{1+x_0}{8}+\frac{\pi^2}{24}(1+x_0)
-{x_0^2\over4(1-x_0)}\ln x_0-{x_0^2\over4(1-x_0)}\ln^2x_0\right]
+{\cal O}(\epsilon^2)\right\},\nonumber\\
\delta_Z^{\rm irr(1)}&=&\frac{N_cg_0^2}{16\pi^2c_0^2M_{Z0}^2}
\sum_{F=A,B}m_{F0}^2\left(\frac{\mu}{m_{F0}}\right)^{2\epsilon}
\left[\frac{1}{2\epsilon}-\frac{1}{2}+\frac{\pi^2}{24}\epsilon
+{\cal O}(\epsilon^2)\right].
\label{delvirr}
\end{eqnarray}
Inserting Eqs.~(\ref{azz1},\ref{aww1},\ref{delvirr}) into Eq.~(\ref{delv}), we
obtain the final one-loop results as
\begin{eqnarray}
\hat\delta_W^{(1)}&=&-\frac{N_c\hat g^2m_A^2}{16\pi^2M_W^2}\,\frac{1+x}{2},
\nonumber\\
\hat\delta_Z^{(1)}&=&-\frac{N_c\hat g^2m_A^2}{16\pi^2\hat c^2M_Z^2}\,
\frac{1+x}{2}.
\label{wz1}
\end{eqnarray}

Expanding Eq.~(\ref{delv}) and collecting the two-loop pieces, we have
\begin{equation}
\hat\delta_V^{(2)}=\delta_V^{\rm irr(2)}-\frac{\delta^{(2)}M_V^2}{M_V^2}
+\delta_V^{\rm ct}-\delta_V^{\rm irr(1)}\frac{\delta^{(1)}M_V^2}{M_V^2}.
\label{hww2}
\end{equation}
As in case of $\delta_u^{\rm ct}$, the counterterm contribution,
$\delta_V^{\rm ct}$, consists of two parts which scale as $N_c$ and $N_c^2$.
They emerge by renormalizing the heavy-fermion and intermediate-boson masses,
respectively, in the bare analogue of Eq.~(\ref{wz1}),
$\hat\delta_V^{(1)}=\delta_V^{\rm irr(1)}-\delta^{(1)}M_V^2/M_V^2$, where
$\delta_V^{\rm irr(1)}$ is given in Eq.~(\ref{delvirr}) and
$\delta^{(1)}M_V^2/M_V^2$ in Eqs.~(\ref{azz1},\ref{aww1}).
The part proportional to $N_c^2$ exactly cancels against the last term on the
right-hand side of Eq.~(\ref{hww2}).
The part proportional to $N_c$ is obtained by shifting the fermion masses in
the bare version of $\hat\delta_V^{(1)}$ according to Eq.~(\ref{quarkct}) and
reads
\begin{eqnarray}
\delta_W^{\rm ct}&\doteq&-{3(1-x)^2\over8\epsilon}-{15-26x+15x^2\over16}
+{x^2(8-x-x^2)\over8(1-x)}\ln x-{(1-x)^4\over8x}\ln|1-x|
+{\cal O}(\epsilon),\nonumber\\
\delta_Z^{\rm ct}&\doteq&-{3(1-x)^2\over8\epsilon}-{3-2x+3x^2\over4}
-{x\over8}(3-2x-x^2)\ln x-{(1-x)^4\over8x}\ln|1-x|
+{\cal O}(\epsilon).\nonumber\\
\end{eqnarray}
Again, the knowledge of the ${\cal O}(\epsilon)$ term in Eq.~(\ref{quarkct})
is not required, as Eq.~(\ref{wz1}) does not contain ultraviolet divergences.

When we apply Eq.~(\ref{let}), we must differentiate with respect to the
fermion masses which enter through the propagators and treat those coming 
from the Yukawa couplings as constants.
We thus obtain
\begin{eqnarray}
\delta_W^{\rm irr(2)}&\doteq&-{3(1-x)^2\over16\epsilon^2}
+\frac{1}{\epsilon}\left[{9-10x+9x^2\over32}
+{3x^2(3-x)\over8(1-x)}\ln x\right]
\nonumber\\
&&+{3\over64}(27-14x+27x^2)-{\pi^2\over32}(1-x)^2
+{x(14-5x+9x^2)\over16(1-x)}\ln x
\nonumber\\ 
&&-{x^2(7-10x-x^2)\over16(1-x)^2}\ln^2x-{7\over8}(1-x^2)\di(1-x)
+{\cal O}(\epsilon),
\nonumber\\
\delta_Z^{\rm irr(2)}&\doteq&-{3(1-x)^2\over16\epsilon^2}
+\frac{1}{\epsilon}\left[{15-46x+15x^2\over32}-{3\over8}x(1-x)\ln x\right]
\nonumber\\
&&+{7+50x+7x^2\over64}-{\pi^2\over32}(1-x)^2+{x\over16}(23-15x)\ln x
\nonumber\\
&&+{x\over16}(3-5x)\ln^2x-{1-x^2\over8}\di(1-x)+{\cal O}(\epsilon).
\end{eqnarray}

Combining the previous results, we finally obtain for the complete two-loop
correction to the $HWW$ and $HZZ$ couplings in the $\msbar$ scheme
\begin{eqnarray}
\hat\delta_W^{(2)}&\doteq&-{3\over4}(1-x+x^2)-{x^2(10-5x+x^2)\over8(1-x)}\ln x
-{(1-x)^4\over8x}\ln|1-x|,
\nonumber\\
\hat\delta_Z^{(2)}&\doteq&-{15\over16}(1-x)^2+{x\over8}(3-4x+x^2)\ln x
-{(1-x)^4\over8x}\ln|1-x|,
\end{eqnarray}
respectively.
For $m_A\gg m_B$ and $m_A\approx m_B$, this reduces to
\begin{eqnarray}
\left.\hat\delta_W^{(2)}\right|_{x=0}&\doteq&-\frac{5}{8},\qquad
\left.\hat\delta_Z^{(2)}\right|_{x=0}\doteq-\frac{13}{16},\nonumber\\
\left.\hat\delta_W^{(2)}\right|_{x=1}&=&
\left.\hat\delta_Z^{(2)}\right|_{x=1}=0,
\end{eqnarray}
respectively.

\subsection*{Results in the on-shell scheme}

The difference between the contributions to the $HVV$ vertices in the 
electroweak $\msbar$ and on-shell schemes is not only due to the universal
part discussed in the Section~2, but also due to terms specific to the
$HVV$ interactions.
To elucidate this point, we rewrite Eq.~(\ref{lagrHVV}) in the on-shell scheme
as
\begin{equation}
{\cal L}_{HVV}=2^{1/4}G_\mu^{1/2}H\left(1+\delta_u\right)
\left[2M_W^2W_\mu^\dag W^\mu(1+\delta_W)+M_Z^2Z_\mu Z^\mu(1+\delta_Z)\right].
\label{lagrz}
\end{equation}
In the case of the $HWW$ coupling, it follows from
Eqs.~(\ref{replace},\ref{wz1}) that we may identify
\begin{equation}
\delta_W^{(1)}=\hat{\delta}_W^{(1)},\qquad
\delta_W^{(2)}=\hat{\delta}_W^{(2)},
\end{equation}
where it is understood that the couplings on the right-hand sides are to be
expressed in terms of $G_\mu$.

On the other hand, in the case of the $HZZ$ coupling, one needs to take into
account the shift in $\hat c^2$. 
At the one-loop order, this does not lead to any change, so that
\begin{equation}
\delta_Z^{(1)}=\hat\delta_Z^{(1)}.
\end{equation}
At two loops, however, an extra contribution arises from the denominator of
Eq.~(\ref{wz1}), so that
\begin{eqnarray}
\delta_Z^{(2)}&=&\hat\delta_Z^{(2)}+\delta_Z^{(1)}\Delta\rho^{(1)}\nonumber\\
&\doteq&-\frac{15}{16}(1-x)^2+\frac{x}{8}(3-4x+x^2)\ln x
-\frac{(1-x)^4}{8x}\ln|1-x|\nonumber\\
&&-\frac{N_c}{8}(1+x)\left[1+x+\frac{2x}{1-x}\ln x\right],
\end{eqnarray} 
with the limits
\begin{eqnarray}
\left.\delta_Z^{(2)}\right|_{x=0}\doteq-\frac{13}{16}-\frac{N_c}{8},\qquad
\left.\delta_Z^{(2)}\right|_{x=1}=0.
\end{eqnarray}

Finally, we can combine the universal and vertex-specific corrections to
obtain the complete one- and two-loop corrections to the Lagrangian of
Eq.~(\ref{lagrz}) as
\begin{eqnarray}
\delta_V^{\rm tot(1)}&=&\delta_u^{(1)}+\delta_V^{(1)},\nonumber\\
\delta_V^{\rm tot(2)}&=&\delta_u^{(2)}+\delta_V^{(2)}
+\delta_u^{(1)}\delta_V^{(1)},
\end{eqnarray}
respectively.
At one loop, we recover the results of Ref.~\cite{cha,hvv}.
This completes the discussion of the $HWW$ and $HZZ$ couplings.

\boldmath
\section{Loop-induced interactions}
\unboldmath

The $H\gamma\gamma$ coupling is mediated via loops of $W$ bosons and charged
fermions, while the $Hgg$ coupling is generated by quark loops (see Fig.~3a).
These virtual particles do not decouple if their masses are much larger than
$M_H$, since their couplings to the Higgs boson grow with their masses.
Therefore, the $H\gamma\gamma$ and $Hgg$ vertices provide devices to count the
number of charged and coloured high-mass particles, respectively, which may be
too heavy to be produced on their mass shells.
Owing to QED-like Ward identities, the leading two-loop electroweak
corrections to the $H\gamma\gamma$ and $Hgg$ couplings scale only
quadratically with the heavy-fermion masses, i.e.\ they are of
${\cal O}(G_\mu m_F^2)$ relative to the Born approximation.
In the next subsection, we will derive the ${\cal O}(G_\mu m_F^2)$ correction
to the $H\gamma\gamma$ amplitude.
The corresponding result for the $Hgg$ coupling can then be readily obtained
from this by adjusting the coupling constants and colour factors.

\boldmath
\subsection{The $H\gamma\gamma$ coupling}
\unboldmath

For simplicity, we will assume that all three external legs are on their mass
shells and that the mass hierarchy $M_H/2\ll M_W\ll m_F$ holds.
Thus, our results should be valid for an intermediate-mass Higgs boson.
By virtue of electromagnetic gauge invariance, the $H\gamma\gamma$ amplitude
must be proportional to $(k_1\cdot k_2\,g^{\mu\nu}-k_2^\mu k_1^\nu)$, where
$k_1$, $k_2$ and $\mu$, $\nu$ are the four-momenta and Lorentz indices of the
two photons, respectively.
We first consider the one-loop contribution due to a weak isodoublet of
ultraheavy fermions, with masses $m_A$, $m_B$, electric charges $Q_A$, $Q_B$
and colour multiplicity $N_c$.
At one loop, the effective Lagrangian describing the resulting $H\gamma\gamma$
interaction reads in bare form \cite{djo,let2}
\begin{equation}
{\cal L}_{H\gamma\gamma}=\frac{e_0^2}{24\pi^2}\,\frac{H_0}{v_0}
F_{0\mu\nu}F_0^{\mu\nu}\sum_{F=A,B}N_cQ_F^2,
\label{LHgg1}
\end{equation}
where $e$ is the electromagnetic coupling constant and $F_{\mu\nu}$ is the
electromagnetic field-strength tensor.
Up to terms of ${\cal O}(\epsilon)$, the corresponding coefficient of
$(k_1\cdot k_2\,g^{\mu\nu}-k_2^\mu k_1^\nu)$ is thus given by
\begin{equation}
A_F^{(1)}=-\frac{8}{3v_0}\,\frac{e_0^2}{16\pi^2}\sum_{F=A,B}N_cQ_F^2
\left(\frac{\mu}{m_{F0}}\right)^{2\epsilon}[1+{\cal O}(\epsilon^2)]. 
\label{hggferm1}
\end{equation}
The contributions due to light fermions are negligible.

The residual one-loop contributions, due to $W$ bosons, charged
pseudo-Goldstone bosons ($\phi$) and Faddeev-Popov ghosts, are conveniently
calculated in a non-linear gauge \cite{fujikawa,fawzi}, which has frequently
been used to simplify the calculation of loop amplitudes involving external
photons \cite{gavela}.
An equivalent alternative makes use of the background-field method
\cite{russians,BFM}.
In the non-linear gauge, the $\gamma W^\pm\phi^\mp$ coupling, which is
characteristic for the $R_\xi$ gauge, is avoided, so that the photon 
separately couples to $W$ and charged Goldstone bosons.
As a consequence, the bosonic $H\gamma\gamma$ loop diagrams come in a smaller
number than in the $R_\xi$ gauge, and they can be separated in $W$-boson and
unphysical-scalar-boson parts.
Further modifications \cite{barroso} in the Faddeev-Popov sector of the SM do
not affect the calculation, as ghosts do not couple to fermions.
As the $H\phi^+\phi^-$ coupling is proportional to $M_H$, the diagrams
involving the Goldstone bosons can be discarded in our approximation. 
Keeping the ${\cal O}(\epsilon)$ parts and fixing the gauge-parameter to one,
we have for the two remaining contributions \cite{russians,gavela}:
\begin{eqnarray}
A_W^{(1)}&=&\frac{2}{v_0}\,\frac{e_0^2}{16\pi^2}
\left(\frac{\mu}{M_W^0}\right)^{2\epsilon}
\left[\frac{20}{3}+\frac{2}{3}\epsilon+{\cal O}(\epsilon^2)\right],
\nonumber\\
A_{FP}^{(1)}&=&\frac{2}{v_0}\,\frac{e_0^2}{16\pi^2} 
\left(\frac{\mu}{M_W}\right)^{2\epsilon}
\left[\frac{1}{3}+{\cal O}(\epsilon^2)\right].
\label{bos1}
\end{eqnarray}
The full $H\gamma\gamma$ amplitude to one loop is then obtained as
$A^{(1)}=A_F^{(1)}+A_W^{(1)}+A_{FP}^{(1)}$. 

We now turn to the leading two-loop correction to the $H\gamma\gamma$
amplitude.
One may wonder whether the Yukawa Lagrangian characterizing the gaugeless
limit of the SM \cite{bar}, which we have used in the previous sections to
calculate ${\cal O}(G_\mu^2m_F^4)$ corrections to the $Hf\bar f$ and $HVV$
couplings, is sufficient to also extract the ${\cal O}(G_\mu m_F^2)$
correction to the $H\gamma\gamma$ coupling.
In fact, in contrast to the previous cases, Ward identities insure the absence
of ${\cal O}(G_\mu m_F^4)$ contributions, and it is not obvious any more
that diagrams involving virtual $W$ bosons may be neglected.
This can be understood in the following way.
After integrating out the heavy fermions, we have two dimension-four operators
which produce ${\cal O}(G_\mu m_F^2)$ corrections to the $H\gamma\gamma$
amplitude at the two-loop level, namely the operator in Eq.~(\ref{LHgg1}) and
the first one in Eq.~(\ref{lagrHVV}), through the diagram of Fig.~3b.
Hence, the effective Lagrangian appropriate for the problem at hand is
\begin{equation}
{\cal L}_{H\gamma\gamma}=\frac{H_0}{v_0}\left(
c_1F_{0\mu\nu}F_0^{\mu\nu}+c_2M_{W0}^2W_{0\mu}^\dag W_0^\mu+\ldots\right),
\label{leff}
\end{equation}
where the ellipsis stands for additional operators involving charged
pseudo-Goldstone bosons, such as $\partial_\mu\phi^\dag\partial^\mu\phi$,
which are characteristic for the covariant gauges and are absent in the
unitary gauge; for a detailed discussion of the low-energy effective
Lagrangian of the SM, see e.g.\ Ref.~\cite{fer}.
For the problem at hand, we only need the one-loop result for $c_2$, which is
available from Section~3.
By contrast, $c_1$ must be determined by an explicit two-loop calculation.
As in the previous sections, this calculation can be performed by considering
the diagrams which only involve virtual Goldstone bosons along with the heavy
fermions.
The diagrams containing only virtual heavy fermions and scalar bosons are  
described by a Yukawa Lagrangian in which the would-be Goldstone bosons are
minimally coupled to the photon.
Once the relevant two-loop contribution to the photon self-energy function
has been calculated from this Lagrangian, one can derive the corresponding
$H\gamma\gamma$ vertex correction using a low-energy theorem similar to
Eq.~(\ref{let}).
We have verified that the photon self-energy tensor, supplemented by the
scalar-boson mass counterterms as explained in Section~2, is transverse and
satisfies the QED Ward identity.
Note that $\Pi_{\gamma\gamma}^\prime(0)$, where $\Pi_{\gamma\gamma}(q^2)$ is
the photon self-energy function, contains contributions which depend
logarithmically on the scalar-boson masses $m_\phi$.
Thus, they are infrared divergent in the limit $m_\phi\to0$ and gauge
dependent in the $R_\xi$ gauges.
However, in our approximation of neglecting all scalar-boson couplings to the
Higgs boson, the degree of convergence of the irreducible vertex diagrams is
higher than the one of the photon self-energy, so that the application of the 
low-energy theorem leads to an infrared- and ultraviolet-finite and
gauge-invariant result for the $H\gamma\gamma$ amplitude.\footnote{As 
$\Pi_{\gamma\gamma}^\prime(0)$ is infrared divergent even after subtracting
the scalar-boson mass counterterms, particular care must be exercised in
regularizing these divergences.
Logarithmic divergences which have the form $\ln m_\phi$ if we introduce a
small mass regulator $m_\phi$, turn into $1/\epsilon - \ln m_F$ if we use
dimensional regularization. 
The additional $\ln m_F$ terms introduced in this way conflict with the use of
the low-energy theorem~(\ref{let}), which leads one to also differentiate such
terms with respect to $m_F$.
Therefore, dimensional regularization of the infrared divergences should be
avoided here.}

Let us first consider the diagrams in the first row of Fig.~2b, in which both
photons couple to the heavy fermions.
With the help of the low-energy theorem~(\ref{let}), their contribution to the
irreducible $H\gamma\gamma$ vertex is found to be
\begin{eqnarray}
A_\phi^{(2)I}&\doteq&\frac{1+x}{3\epsilon}(Q_A-Q_B)^2
-\frac{1}{9}(4Q_A^2+Q_AQ_B-23Q_B^2)\nonumber\\
&&+\frac{x}{9}(23Q_A^2-Q_AQ_B-4Q_B^2)+\frac{2}{3}(1+x)Q_B(Q_A-Q_B)\ln x.
\label{4point}
\end{eqnarray}
Here and henceforth, the symbol $\doteq$ indicates that the right-hand side is
to be multiplied by the overall factor
\begin{equation}
\frac{N_ce^2\hat g^3m_A^2}{(16\pi^2)^2M_W^3}  
\left(\frac{\mu}{m_A}\right)^{4\epsilon}.
\label{units}
\end{equation} 
The electric charges of the isodoublet partners satisfy the relation
$Q_A-Q_B=1$.
We note in passing that, if $Q_A=Q_B$ was fulfilled, Eq.~(\ref{4point}) would
be ultraviolet finite and free of singularities in the limit of one fermion
mass vanishing.
In fact, as we will see in the next subsection, this corresponds to the $Hgg$
case.

Next, we consider the diagrams in the second row of Fig.~2b, where the photons
couple at least to one charged scalar boson.
Also including the contributions due to the scalar-boson mass counterterms, we
find
\begin{eqnarray}
A_\phi^{(2)II}&\doteq&-{1+x\over3\epsilon}(Q_A-Q_B)
+{(1+x)\over18}(1-5Q_A+5Q_B)
\nonumber\\
&&-\frac{\ln x}{3(1-x)}[2Q_B+(1-Q_A-Q_B)x^2]
-\frac{1+x}{3}(1-Q_A+Q_B)\ln\frac{m_\phi^2}{m_A^2}.
\label{aphiII}
\end{eqnarray}
Due to $Q_B=Q_A-1$, this is convergent as $m_\phi\to0$ and cancels the
ultraviolet divergence of Eq.~(\ref{4point}).
We still need to take into account the counterterm contribution which arises
by shifting the bare heavy-fermion masses in Eq.~(\ref{hggferm1}) according to
Eq.~(\ref{quarkct}), which reads
\begin{equation}
A_\phi^{\rm ct}\doteq-(1-x)(Q_A^2-Q_B^2).
\label{2loopct}
\end{equation}
We recall that the renormalization of $e$ does not generate any
${\cal O}(G_\mu m_F^2)$ corrections.
Adding Eqs.~(\ref{4point},\ref{aphiII},\ref{2loopct}), we obtain the final
result for the scalar-boson exchange diagrams,
\begin{equation}
A_\phi^{(2)}\doteq\frac{10}{3}-7Q_A+2Q_A^2
-x\left(\frac{5}{3}-3Q_A-2Q_A^2\right),
\label{2loops}
\end{equation}
which is finite and gauge invariant.

Let us now turn to the diagrams involving $W$ bosons.
With our choice of gauge, we can separately consider the diagrams which,
in addition to the heavy fermions, contain $W$ and charged Goldstone bosons
because the $\gamma W^\pm\phi^\mp$ coupling is absent.
The $W$ and charged Goldstone bosons still mix through the $HW^\pm\phi^\mp$
coupling, but the leading contribution from the two-loop $H\gamma\gamma$
diagrams which involve this coupling can be removed by an appropriate choice
of renormalization in the unphysical sector \cite{renorm},
which respects the Slavnov-Taylor identities \cite{FTJ}.
The same choice also insures that $M_W$ in the expression for $A_{FP}$ in
Eq.~(\ref{bos1}) does not need to be renormalized.
The only diagrams containing virtual $W$ bosons which can generate
${\cal O}(G_\mu m_F^2)$ corrections are those with a quadratic subdivergence.
This includes the diagrams where a fermion loop is inserted into a $W$-boson
line.
In this case, however, the quadratic subdivergences are cancelled by the
corresponding $W$-boson mass counterterms.
We are then just left with the contribution from the left diagram in Fig.~3b,
which corresponds to the insertion of the second operator in the effective
Lagrangian~(\ref{leff}) into the one-loop seed diagram containing a $W$-boson
loop.
This contribution can be easily calculated from Eqs.~(\ref{wz1},\ref{bos1})
and reads
\begin{equation}
A_W^{(2)}\doteq-\frac{10}{3}(1+x).
\label{2loopw}
\end{equation}
This includes the $W$-boson mass-counterterm contribution that arises from
$A_W^{(1)}$, and has been confirmed by an explicit two-loop calculation.

Finally, we need to include the finite contribution due to the shift in the
ratio $H_0/v_0$, i.e.\ $\hat\delta_u^{(1)}$ of Eq.~(\ref{ms1}) in the 
electroweak $\msbar$ scheme and $\delta_u^{(1)}$ of Eq.~(\ref{os1}) in the
on-shell scheme.
This is the only scheme-dependent contribution at the two-loop level.
As $H_0/v_0$ is an overall factor common to all one-loop contributions,
the shift operates on the total one-loop amplitude, 
giving the contributions
\begin{eqnarray}
\hat A_u^{(2)}&\doteq&\frac{2}{3}(1+x)
\left(\frac{7}{4}-\frac{1}{3}\sum_FN_cQ_F^2\right),
\nonumber\\
A_u^{(2)}&\doteq&\left[\frac{7}{6}(1+x)+\frac{x}{1-x}\ln x\right]
\left(\frac{7}{4}-\frac{1}{3}\sum_FN_cQ_F^2\right)
\label{shiftu}
\end{eqnarray}
in the $\msbar$ and on-shell schemes, respectively.

The total ${\cal O}(G_\mu m_F^2)$ correction $A^{(2)}$ to the $H\gamma\gamma$
amplitude is then obtained by summing the contributions of
Eqs.~(\ref{2loops}--\ref{shiftu}).
In the on-shell scheme, the result reads
\begin{equation}
A^{(2)}\doteq A_u^{(2)}-Q_A(7-2Q_A)-x(1-Q_A)(5+2Q_A).
\end{equation}
In the limiting cases $m_A\gg m_B$ and $m_A\approx m_B$, this reduces to
\begin{eqnarray}
\label{azero}
\left.A^{(2)}\right|_{x=0}&\doteq&{49\over24}-7Q_A+2Q_A^2
-\frac{7}{18}N_c(1-2Q_A+2Q_A^2),\\
\left.A^{(2)}\right|_{x=1}&\doteq&-{8\over3}-4Q_A+4Q_A^2
-\frac{4}{9}N_c(1-2Q_A+2Q_A^2),
\end{eqnarray}
respectively.
Notice, however, that Eq.~(\ref{azero}) is only valid in the case of two
ultraheavy fermions with a strong mass hierarchy, i.e.\ $M_W\ll m_B\ll m_A$,
and thus cannot be applied to the contribution from the top and bottom quarks.
In the latter case, $m_b\ll M_W\ll m_t$, the bottom quark decouples from the
one-loop result, so that Eqs.~(\ref{2loopct},\ref{shiftu}) must be modified.
Then, Eq.~(\ref{azero}) becomes
\begin{equation}
A_{tb}^{(2)}\doteq\frac{25}{24}-5Q_t+Q_t^2-\frac{7}{18}N_cQ_t^2.
\end{equation}

In the three-generation SM with $M_H/2\ll M_W\ll m_t$, the
${\cal O}(G_\mu m_t^2)$-corrected $H\gamma\gamma$ amplitude, renormalized 
according to the on-shell scheme, reads
\begin{equation}
A_{\rm SM}=\frac{\alpha G_\mu^{1/2}}{\pi2^{3/4}}\,\frac{47}{9}
\left[1+\frac{G_\mu}{8\pi^2\sqrt2}\left(-\frac{511}{94}m_t^2\right)\right],
\label{asm}
\end{equation}
where $\alpha=e^2/(4\pi)$ is Sommerfeld's fine-structure constant.
Large logarithmic QED corrections can be avoided by taking $\alpha$ to be the
running coupling evaluated at a renormalization scale of the order of the 
heavy-quark mass.
Should the SM be extended by a sequential fermion generation consisting of a
Dirac neutrino $N$, a charged lepton $E$, an up-quark $U$ and a down-quark
$D$, with masses $m_N,m_E,m_U,m_D\gg M_W$, then their effects on $A^{(1)}$,
$A^{(2)}$ and $\delta_u^{(1)}$ would come in addition to the respective
top-quark contributions.
Thus, Eq.~(\ref{asm}) would be modified to become
\begin{eqnarray}
A_{\rm 4gen}&=&\frac{\alpha G_\mu^{1/2}}{\pi2^{3/4}}\,\frac{5}{3}\left[1
+\frac{G_\mu}{8\pi^2\sqrt2}\left(-\frac{49}{2}m_t^2
+\frac{7}{6}m_N^2-\frac{65}{6}m_E^2
-\frac{m_N^2m_E^2}{m_N^2-m_E^2}\ln\frac{m_N^2}{m_E^2}
\right.\right.\nonumber\\
&&-\left.\left.
\frac{237}{10}m_U^2-\frac{117}{10}m_D^2
-3\frac{m_U^2m_D^2}{m_U^2-m_D^2}\ln\frac{m_U^2}{m_D^2}\right)\right].
\label{a4g}
\end{eqnarray}

All previous results for the $H\gamma\gamma$ amplitude were derived under the
assumption that $M_H\ll2M_W$.
A convenient way to perform the calculation without this restriction would be
to employ an effective-Lagrangian approach, and to calculate the contributions
of the various diagrams as insertions of appropriate operators whose
coefficient functions incorporate the low-energy effect of the heavy fermions.
We recall that, in a covariant gauge, the Lagrangian~(\ref{leff}) is
supplemented by additional operators involving charged pseudo-Goldstone bosons.
Similarly to the use of the second operator in Eq.~(\ref{leff}) for the
diagram of Fig.~3b, they will only enter the calculation through insertions in
one-loop diagrams.
Therefore, their Wilson coefficients are only needed to
${\cal O}(G_\mu m_F^2)$.
The only operator whose matching is needed at the two-loop order is the first
one in Eq.~(\ref{leff}).
As only the diagrams in the first row of Fig.~2b do not have a counterpart in
the effective theory, this matching is uniquely fixed by
Eqs.~(\ref{4point},\ref{2loopct}), which exhaust all necessary genuine
two-loop calculations.
If the restriction $M_H\ll2M_W$ is abandoned, $A_W^{(1)}$, $A_{FP}^{(1)}$ and
the corresponding amplitude for the charged pseudo-Goldstone bosons will be
multiplied by non-analytic functions of $M_H/M_W$ whose limits for $M_H\to0$
are either unity or zero; see, for example, Refs.~\cite{kni,russians}. 

In this paper, we do not consider the ${\cal O}(G_\mu^2 m_F^4)$ correction to
the $H\gamma Z$ coupling.
At one loop, the fermionic contribution to the $H\gamma Z$ amplitude is
greatly suppressed against the $W$-boson contribution, and it is likely that
this trend carries over to the two-loop order.
Furthermore, from the phenomenological point of view, this coupling is much
less important than the $H\gamma\gamma$ and $Hgg$ couplings.
The ${\cal O}(G_\mu^2 m_F^4)$ correction to the $H\gamma Z$ coupling could be
calculated along the same lines as those for the $H\gamma\gamma$ coupling, but
choosing a different kind of non-linear gauge.
In fact, it is possible to fix the gauge in such a way that the
$ZW^\pm\phi^\mp$ vertex is avoided, so that the $Z$ boson separately couples
to the $W$ and charged Goldstone bosons separately.
As in the $H\gamma\gamma$ case, modifications in the Faddeev-Popov sector
would not affect the calculation.

\boldmath
\subsection{The $Hgg$ coupling}
\unboldmath

We now study the two-loop ${\cal O}(G_\mu m_F^2)$ corrections to the $Hgg$
coupling.
They can be inferred from those $H\gamma\gamma$ diagrams where the photons are
directly coupled to the loop fermions.
The relevant amplitudes are those given by
Eqs.~(\ref{hggferm1},\ref{4point},\ref{2loopct}) as well as the term
proportional to $N_c$ in Eq.~(\ref{shiftu}).
We need to identify $Q_A=Q_B=1$ and substitute $\alpha\to\alpha_s$.
Furthermore, we need to adjust the overall colour factor by putting
$N_c\to\delta^{ab}/2$ in Eqs.~(\ref{hggferm1},\ref{units}).
Here, $a$ and $b$ are the colour indices of the two gluons.
In the following, we shall factor out $\delta^{ab}$ along with
$(k_1\cdot k_2\,g^{\mu\nu}-k_2^\mu k_1^\nu)$.

In the three-generation SM with $M_H\ll2m_t$, the
${\cal O}(G_\mu m_t^2)$-corrected $Hgg$ amplitude, renormalized according to
the electroweak on-shell scheme, is thus found to be
\begin{equation}
G_{\rm SM}=-\frac{2^{1/4}\alpha_sG_\mu^{1/2}}{3\pi}\left[1
+\frac{G_\mu m_t^2}{8\pi^2\sqrt2}\left(\frac{7}{6}N_c-3\right)\right],
\label{gsm}
\end{equation}
where $N_c=3$, in agreement with Ref.~\cite{djo}.
If the SM was extended by a sequential fermion generation consisting of a
Dirac neutrino $N$, a charged lepton $E$, an up-quark $U$ and a down-quark
$D$, with masses $m_N,m_E,m_U,m_D\gg M_H/2$, then Eq.~(\ref{gsm}) would become
\begin{eqnarray}
G_{\rm4gen}&=&-\frac{2^{1/4}\alpha_sG_\mu^{1/2}}{\pi}\left\{1
+\frac{G_\mu}{8\pi^2\sqrt2}\left[
\left(\frac{7}{6}N_c-1\right)m_t^2
+\frac{7}{6}(m_N^2+m_E^2)-\frac{m_N^2m_E^2}{m_N^2-m_E^2}\ln\frac{m_N^2}{m_E^2}
\right.\right.\nonumber\\
&&+\left.\left.
\left(\frac{7}{6}N_c-2\right)(m_U^2+m_D^2)
-N_c\frac{m_U^2m_D^2}{m_U^2-m_D^2}\ln\frac{m_U^2}{m_D^2}\right]\right\}.
\label{g4g}
\end{eqnarray}
Here, we have also assumed that $M_H\ll2m_t$, so that the lowest-order
amplitude receives equal contributions from the top, $U$ and $D$ quarks.
The heavy leptons $N$ and $E$ only contribute through the renormalizations of
the Higgs-boson wave-function and the vacuum expectation value.
It is amusing to observe that a mass-degenerate heavy-quark isodoublet does
not generate any dominant two-loop correction to the $Hgg$ amplitude.
If we eliminate the top-quark and heavy-lepton contributions from
Eq.~(\ref{g4g}) so as to isolate the contribution due to a heavy-quark
isodoublet, then we recover the corresponding result of Ref.~\cite{djo}.

\section{Discussion}

We considered the possible extension of the SM by a sequential heavy-fermion
generation and analyzed, at two loops in electroweak perturbation theory, the
resulting shifts in the couplings of a light Higgs boson to the SM leptons,
light quarks, intermediate bosons, photons and gluons.
In the cases of the tree-level and loop-induced couplings, these corrections
are of relative orders ${\cal O}(G_\mu^2m_F^4)$ and ${\cal O}(G_\mu m_F^2)$,
respectively, where $m_F$ is a generic heavy-fermion mass, with
$m_F\gg M_W,M_H$.
In the $H\gamma\gamma$ case, we also assumed that $M_H\ll2M_W$.
We obtained the corresponding top-quark-induced corrections of the
three-generation SM as special cases.

In the following, we analyze the numerical significance of our results, in the
electroweak on-shell scheme.
To ${\cal O}(G_\mu m_F^2)$ and ${\cal O}(G_\mu^2m_F^4)$, the shifts in the
$Hf\bar f$, $HWW$ and $HZZ$ couplings due to an isodoublet of heavy fermions,
$A$ and $B$, can all be cast into the generic form
\begin{equation}
\delta=N_c\frac{G_\mu m_A^2}{8\pi^2\sqrt2}C_1\left(\frac{m_B}{m_A}\right)
\left\{1+\frac{G_\mu m_A^2}{8\pi^2\sqrt2}
\left[C_2\left(\frac{m_B}{m_A}\right)
+N_cC_3\left(\frac{m_B}{m_A}\right)\right]\right\},
\label{ctree}
\end{equation}
where $C_i$ $(i=1,2,3)$ are dimensionless functions of $m_B/m_A$ and $N_c=1$
(3) if $A$ and $B$ are leptons (quarks).
Also $\Delta\rho$ can be written in this form, with $C_3=0$.
Since, in the high-$m_F$ limit, $\Delta\rho$, $\delta_u$, $\delta_W$ and
$\delta_Z$ are symmetric in $m_A$ and $m_B$, we may, without loss of
generality, assume that $m_B/m_A\le1$.
The specific forms of the prefactors in Eq.~(\ref{ctree}) are chosen in such a
way that, in the case of the leading $m_t$-dependent contribution to
$\Delta\rho$, the familiar values $C_1(0)=1$ \cite{ros} and
$C_2(0)=19-2\pi^2\approx-0.739$ \cite{hoog} are recovered.
Relative to $m_t=175.6$~GeV, we have
$N_cG_\mu m_A^2/(8\pi^2\sqrt2)\approx0.966\%\times(m_A/m_t)^2$.

The outcome of this decomposition is displayed in Table~1, where the
coefficients $C_i$ are listed as functions of $m_B/m_A$ for $\Delta\rho$,
$\delta_u$, $\delta_W^{\rm tot}$ and $\delta_Z^{\rm tot}$.
Let us first discuss the signs of the various terms.
The ${\cal O}(G_\mu m_F^2)$ terms are throughout positive for $\Delta\rho$ and
$\delta_u$, while they are negative for $\delta_W^{\rm tot}$ and 
$\delta_Z^{\rm tot}$.
In the case of $\delta_u$, the ${\cal O}(G_\mu^2 m_F^4)$ term always enhances
the one-loop correction, while in the other cases this depends on the mass
ratio and on whether leptons or quarks are considered.
{}From the entries for $m_B/m_A=0$ in Table~1, we read off that the
${\cal O}(G_\mu m_t^2)$ corrections to $\Delta\rho$, $\delta_u$,
$\delta_W^{\rm tot}$ and $\delta_Z^{\rm tot}$ amount to 0.966\%, 1.13\%,
$-0.805\%$ and $-0.805\%$, respectively, and that these values receive
relative corrections of $-0.238\%$, 2.42\%, 3.19\% and 6.67\% due to the
${\cal O}(G_\mu^2m_t^4)$ terms.
On the other hand, a mass-degenerate fourth-generation quark isodoublet, with
masses $m_U=m_D=2m_t$, would not influence the $\rho$ parameter, but induce in
$\delta_u$, $\delta_W^{\rm tot}$ and $\delta_Z^{\rm tot}$ one-loop corrections
of 5.15\%, $-10.3\%$ and $-10.3\%$, respectively, which would be enhanced
in magnitude by 11.6\%, 1.93\% and 1.93\% by the two-loop corrections;
see right-most entries in Table~1.

\begin{table}[ht]
\caption{Coefficients $C_1$ (upper entries), $C_2$ (middle entries) and $C_3$
(lower entries) in Eq.~(\ref{ctree}) as functions of $m_B/m_A$ for
$\Delta\rho$, $\delta_u$, $\delta_W^{\rm tot}$ and $\delta_Z^{\rm tot}$ in the
electroweak on-shell scheme.}
\medskip
\begin{center}
\begin{tabular}{|c|c|c|c|c|c|c|} \hline\hline
$m_B/m_A$ &
$  0$ & $  0.2$ & $  0.4$ & $  0.6$ & $  0.8$ & $  1$ \\
\hline
$\Delta\rho$ &
$  1$ & $  0.772$ & $  0.462$ & $  0.211$ & $  0.053$ & $  0$ \\
&
$ -0.739$ & $  0.050$ & $  1.326$ & $  2.423$ & $  2.916$ & $  3$ \\
&
$  0$ & $  0$ & $  0$ & $  0$ & $  0$ & $  0$ \\
\hline
$\delta_u$ &
$  1.167$ & $  1.079$ & $  1.004$ & $  1.012$ & $  1.120$ & $  1.333$ \\
&
$  3.969$ & $  4.696$ & $  5.028$ & $  4.241$ & $  3.139$ & $  3$ \\
&
$  1.179$ & $  1.123$ & $  1.151$ & $  1.329$ & $  1.628$ & $  2$ \\
\hline
$\delta_W^{\rm tot}$ &
$ -0.833$ & $ -1.001$ & $ -1.316$ & $ -1.708$ & $ -2.160$ & $ -2.667$ \\
&
$  6.444$ & $  4.629$ & $  2.449$ & $  0.633$ & $ -0.767$ & $ -1.5$ \\
&
$  1.150$ & $  1.032$ & $  0.892$ & $  0.824$ & $  0.857$ & $  1$ \\
\hline
$\delta_Z^{\rm tot}$ &
$ -0.833$ & $ -1.001$ & $ -1.316$ & $ -1.708$ & $ -2.160$ & $ -2.667$ \\
&
$ 10.044$ & $  7.748$ & $  4.445$ & $  1.568$ & $ -0.528$ & $ -1.5$ \\
&
$  3.550$ & $  2.636$ & $  1.707$ & $  1.159$ & $  0.937$ & $  1$ \\
\hline\hline
\end{tabular}
\end{center}
\end{table}

The partial widths of the Higgs-boson decays into pairs of SM leptons, quarks
and intermediate bosons receive the overall correction factors $(1+\delta)^2$,
where $\delta$ is of the generic form (\ref{ctree}).
In Refs.~\cite{sea,mat}, it was shown how the corresponding corrections to
more complicated Higgs-boson decay and production processes may be composed
from the elementary building blocks $\Delta\rho$, $\delta_u$,
$\delta_W^{\rm tot}$ and $\delta_Z^{\rm tot}$.

We now quantitatively discuss the $H\gamma\gamma$ and $Hgg$ couplings and how 
they are affected by their ${\cal O}(G_\mu m_F^2)$ corrections.
{}From Eqs.~(\ref{asm}) and (\ref{gsm}), we read off that, in the
three-generation SM, the ${\cal O}(G_\mu m_t^2)$ corrections to these
couplings amount to $-1.75\%$ and 0.161\%, respectively.
In the presence of a sequential high-mass fermion generation, the lowest-order
$H\gamma\gamma$ and $Hgg$ amplitudes would be multiplied by factors of
$15/47\approx0.319$ and $3$, respectively; see Eqs.~(\ref{a4g},\ref{g4g}).
Consequently, the $H\to\gamma\gamma$ ($H\to gg$) partial decay width would be
reduced (amplified) by roughly one order of magnitude.
If $m_N=m_E=m_U=m_D=2m_t$, then these modified lowest-order amplitudes would
receive two-loop corrections of $-71.1\%$ and 2.52\%, respectively.
The fact that, in the new-physics scenario considered here, the
${\cal O}(G_\mu m_F^2)$ correction to the $H\gamma\gamma$ coupling is so 
sizeable may cast some doubts on the validity of electroweak perturbation 
theory in this case.
By the same token, one could place upper mass bounds on the fourth-generation
fermions by requiring that the ${\cal O}(G_\mu m_F^2)$ corrections stay
perturbatively small.
However, one should bare in mind that, due to the negative interference of the
bosonic and fermionic diagrams, the one-loop $H\gamma\gamma$ amplitude is 
substantially reduced by the inclusion of the fourth-generation fermions, so
that it may be misleading to use as a criterion the size of the two-loop
correction relative to the one-loop result.

\bigskip
\centerline{\bf ACKNOWLEDGEMENTS}
\smallskip\noindent
We are grateful to P.A. Grassi for many instructive discussions and 
to G. Degrassi and F. Feruglio for interesting conversations.
One of us (A.D.) thanks the Max-Planck-Institut f\"ur Physik for the kind 
hospitality extended to him during a visit when this paper was finalized.

\newpage

\begin{figure}[htb]
\begin{center}
\mbox{\psfig{figure=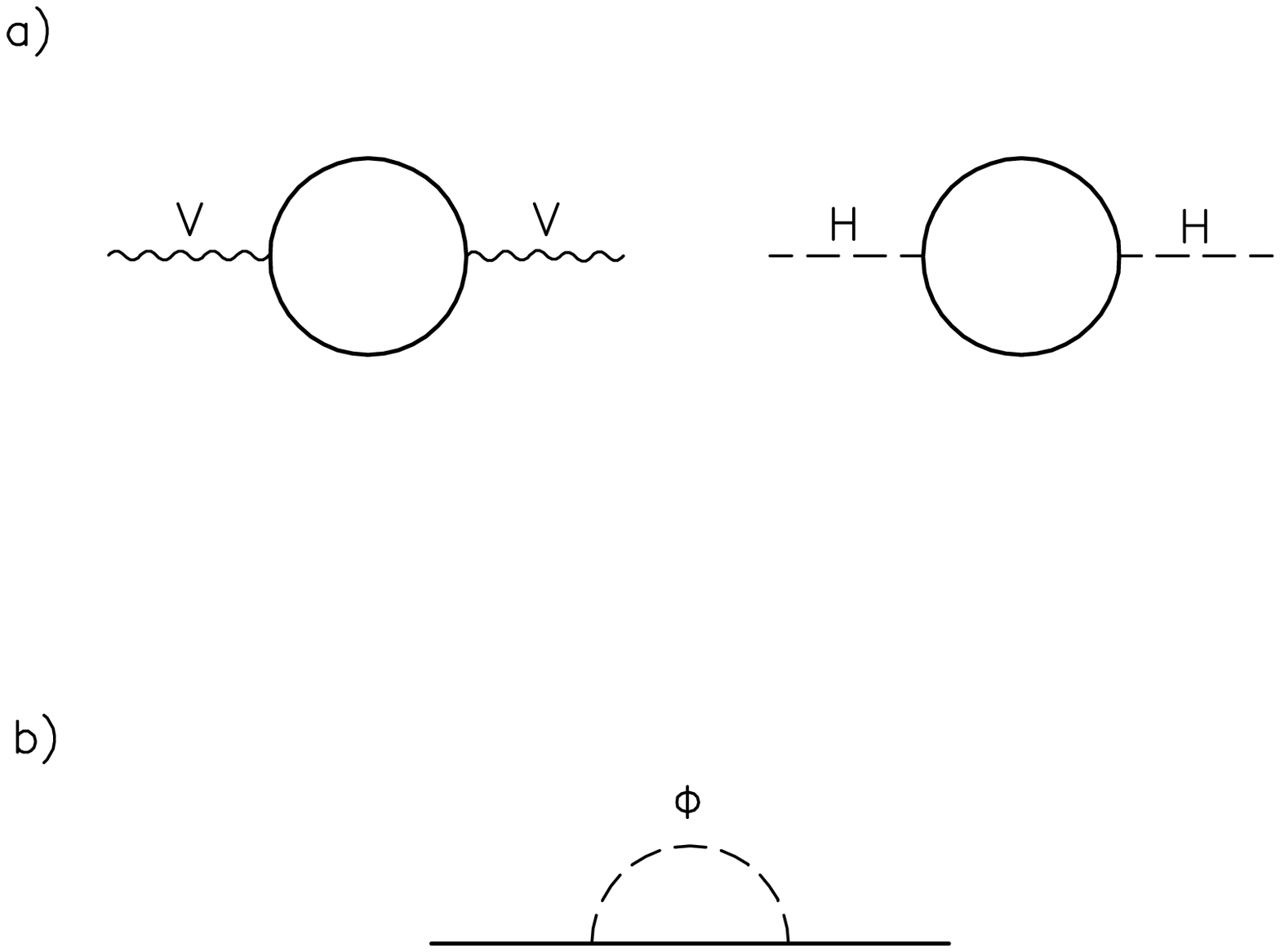,width=12cm}}
\end{center}
\caption[]{One-loop Feynman diagrams generating ${\cal O}(G_\mu m_F^2)$ 
corrections to (a) $\delta^{(1)}M_V^2/M_V^2$, $\Pi_{HH}^{(1)\prime}(0)$ and
(b) $\delta m_F/m_F$.
$\phi$ represents Higgs and pseudo-Goldstone bosons.}
\end{figure}

\begin{figure}[htb]
\begin{center}
\mbox{\psfig{figure=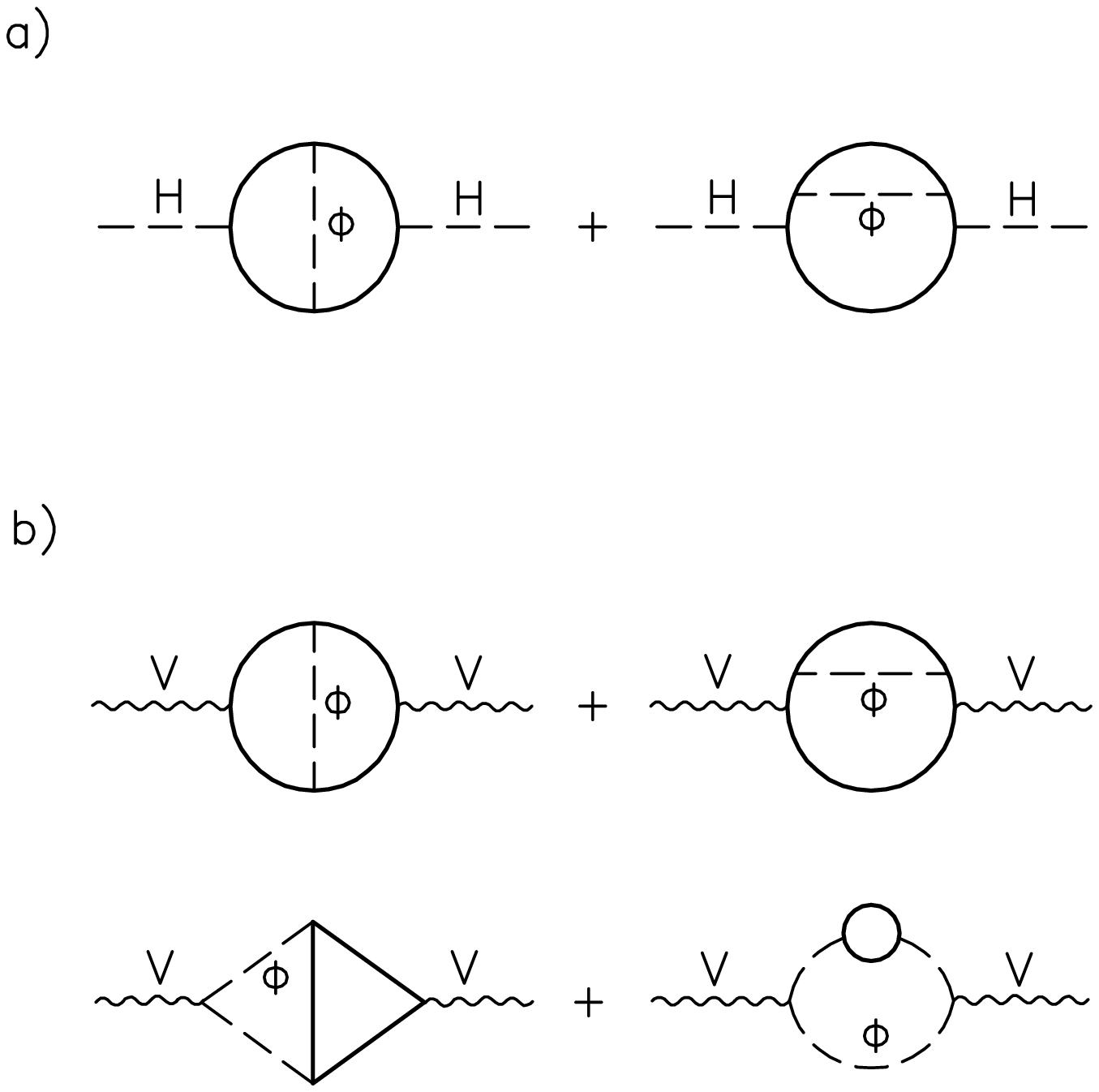,width=12cm}}
\end{center}
\caption[]{Two-loop Feynman diagrams generating ${\cal O}(G_\mu^2m_F^4)$
corrections to (a) $\Pi_{HH}^{(2)\prime}(0)$ and (b)
$\delta^{(2)}M_V^2/M_V^2$.}
\end{figure}

\begin{figure}[htb]
\begin{center}
\mbox{\psfig{figure=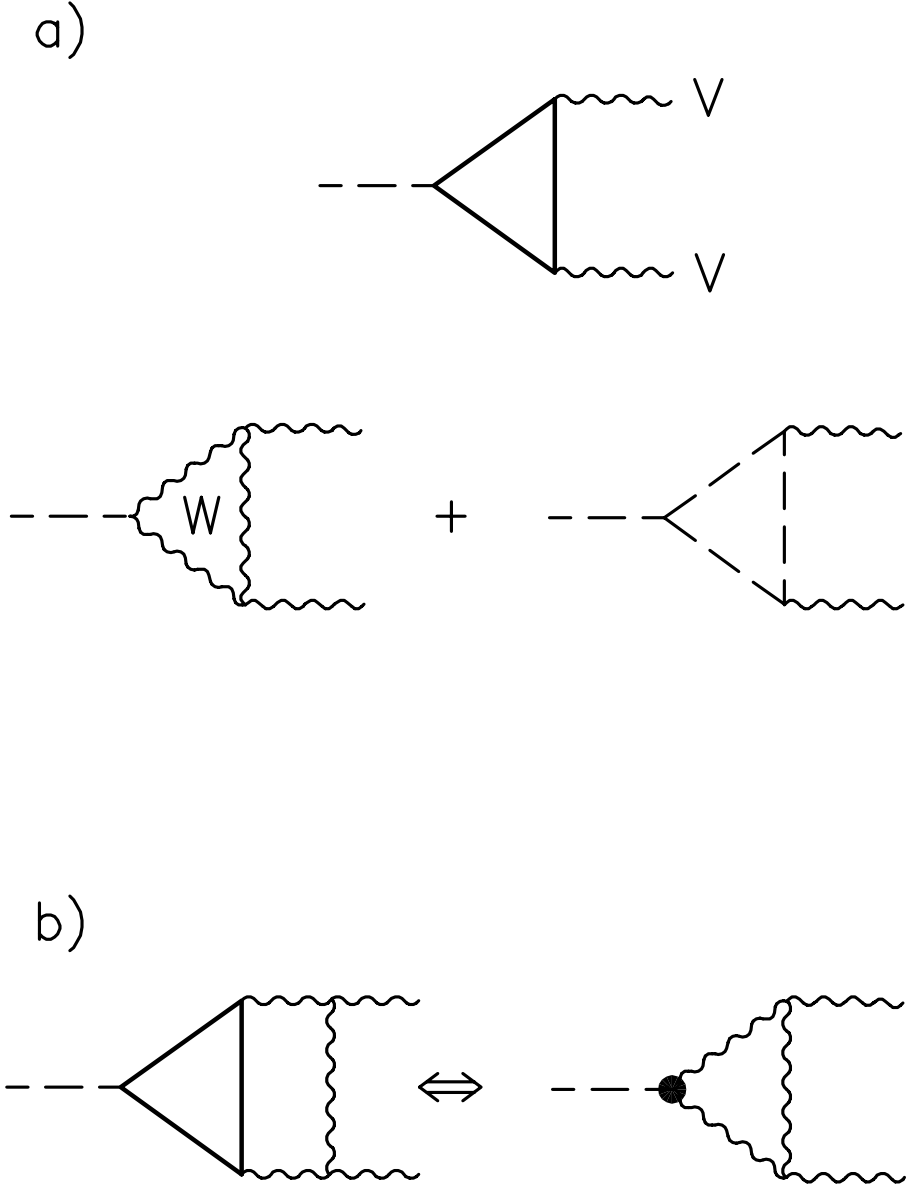,width=12cm}}
\end{center}
\caption[]{(a) One-loop Feynman diagrams generating the leading contributions
$A_F^{(1)}$, $A_W^{(1)}$ and $A_{FP}^{(1)}$ to the $H\gamma\gamma$ amplitude;
(b) two-loop Feynman diagram generating the ${\cal O}(G_\mu m_F^2)$
corrections $A_W^{(2)}$ to the $H\gamma\gamma$ amplitude together with the
equivalent operator insertion.}
\end{figure}

\end{document}